\documentclass[twocolumn,showpacs,preprintnumbers,amsmath,amssymb,superscriptaddress,prb]{revtex4}

\usepackage{graphicx}
\usepackage{dcolumn}
\usepackage{bm}

\begin{document}

\newcommand{\avg}[1]{\langle #1 \rangle}
\newcommand{\var}[1]{\text{Var} (#1)}
\newcommand{\cov}{\text{Cov}}

\title{Effect of Electric Field on Diffusion in Disordered Materials\\
II. Two- and Three-dimensional Hopping Transport}

\author {A. V. Nenashev}
\affiliation{Institute of Semiconductor Physics, 630090 Novosibirsk, Russia}
\affiliation{Novosibirsk State University, 630090 Novosibirsk, Russia}

\author {F. Jansson}
\email {fjansson@abo.fi}
\affiliation{Graduate School of Materials Research,
\AA bo Akademi University,  20500 Turku, Finland}
\affiliation{Department of Physics and Center for Functional Materials,
\AA bo Akademi University, 20500 Turku, Finland}

\author {S. D. Baranovskii}
\affiliation{Department of Physics and Material Sciences Center,
Philipps-University, 35032 Marburg, Germany}

\author {R. \"Osterbacka}
\affiliation{Department of Physics and Center for Functional Materials,
\AA bo Akademi University, 20500 Turku, Finland}

\author {A. V. Dvurechenskii}
\affiliation{Institute of Semiconductor Physics, 630090 Novosibirsk, Russia}
\affiliation{Novosibirsk State University, 630090 Novosibirsk, Russia}

\author{F. Gebhard}
\affiliation{Department of Physics and Material Sciences Center,
Philipps-University, 35032 Marburg, Germany}

\date{\today}

\begin{abstract}
  In the previous paper [Nenashev et al., arXiv:0912.3161] 
  an analytical theory confirmed by numerical
  simulations has been developed for the field-dependent hopping
  diffusion coefficient $D(F)$ in one-dimensional systems with
  Gaussian disorder.  The main result of that paper is the linear,
  non-analytic field dependence of the diffusion coefficient at low
  electric fields.  In the current paper, an  analytical theory is
  developed for the field-dependent diffusion coefficient in three-
  and two-dimensional Gaussian disordered systems in the hopping
  transport regime. The theory predicts a smooth parabolic field
  dependence for the diffusion coefficient at low fields.  The result
  is supported by Monte Carlo computer simulations. In spite of the
  smooth field dependences for the mobility and for the longitudinal
  diffusivity, the traditional Einstein form of the relation between
  these transport coefficients is shown to be violated even at very
  low electric fields.
\end{abstract}




%
\maketitle

\section{Introduction}
\label {sec-introduction}

This paper represents a second part of our research dedicated to
the theory of diffusion of hopping charge carriers biased by electric
field in disordered systems with a Gaussian energy distribution of
the energy of the localized states:
\begin {equation}
\label {eq-gaussian-dos} g (\varepsilon) = \frac {N} {\sigma
\sqrt{2 \pi}} \exp \left( - \frac {\varepsilon^2} {2 \sigma^2}
\right),
\end {equation}
Here $N$ is the spatial concentration of sites available for
hopping transport and $\sigma$ is the energy scale of their
density of states (DOS). The Gaussian DOS is assumed to apply for
disordered organic materials, such as molecularly doped and
conjugated polymers and organic
glasses.\cite{Roth1991,Bassler2000,Pope1999,Bassler1993,
Borsenberger1993,Auweraer1994,Baranovski2006} While in the
previous paper \cite{Nenashev2009I} one-dimensional transport
was considered, in the current paper we present results for two-
and three-dimensional systems. This study has been on one hand
stimulated by numerous experimental studies on organic disordered
materials,\cite{Yuh1988,Borsenberger1991,Borsenberger1993b,Hirao1996,
Hirao1997,Lupton2002,Harada2005} which claim the invalidity of the
conventional form of the Einstein relation between the carrier
mobility $\mu$ and the diffusion coefficient $D$
\begin {equation}
\label {eq-Einstein} \mu = \frac{e}{kT} D.
\end {equation}
On the other hand our study is stimulated by the lack of 
a concise theory for the diffusion biased by
electric field in the hopping transport mode. It is, however, this
transport mode that dominates the electrical conduction in disordered
organic materials where transport is due to incoherent tunnelling
of electrons and holes between localized states randomly
distributed in space, with the DOS described by
Eq.~(\ref{eq-gaussian-dos}).\cite{Roth1991,Bassler2000,Pope1999,
Bassler1993,Borsenberger1993,Auweraer1994} The transition rate
between an occupied state $i$ and an empty state $j$, separated by
the distance $r_{ij}$, is described by the Miller-Abrahams
expression \cite{Miller1960}
\begin {equation}
\label{eq-Miller-Abrahams} \Gamma_{ij} = \nu_0 \,
e^{-2\frac{r_{ij}}{a}} \left\{
\begin{array}{ll}
 e^{ - \frac{\Delta \varepsilon_{ij}}{kT}}
 \qquad & ,\ \Delta \varepsilon_{ij} > 0\\
 1  & ,\ \Delta \varepsilon_{ij} \leq 0
  \end{array}  \right. ,
\end {equation}
where $\nu_0$ is the attempt-to-escape frequency. The energy
difference between the sites is
\begin {equation}
\label{eq-delta} \Delta \varepsilon_{ij} = \varepsilon_j -
\varepsilon_i - F e (z_j - z_i),
\end {equation}
where the electric field $F$ is assumed to be directed along the
$Z$-direction.  The localization length of the charge carriers in
the states contributing to the hopping transport is $a$. We assume the
latter quantity to be independent of energy and we will neglect
correlations between the energies of the localized states,
following the Gaussian-disorder-model of B\"assler. \cite
{Bassler1993,Borsenberger1993,Auweraer1994,Baranovski2006}

The field-dependent diffusion in such systems in the 3D case has so
far been studied by computer simulations. Richert, Pautmeier, and
B\"assler performed Monte Carlo simulations in 3D and showed that the
longitudinal diffusion coefficient $D_z$ is strongly dependent on the
electric field and that the dependence is quadratic at such low fields
that the mobility of charge carriers remains
field-independent.\cite{Pautmeier1991,Richert1989} This is in contrast
to the linear field dependence of the diffusion coefficient at low
fields obtained by the exact analytical theory and by numerical
calculations for 1D systems in the preceding
paper.\cite{Nenashev2009I} In order to clarify the nature of the field
effect on the diffusion for 3D and 2D systems we suggest in the
current paper an analytical theory for the field-dependent diffusion
coefficient in the hopping regime for such systems.  This theory
confirms the conclusion of Richert, Pautmeier, and
B\"assler\cite{Richert1989,Pautmeier1991} about the parabolic field
dependence of $D_z$ at low fields. Furthermore, our theory gives
explicit analytical expressions for the combined effects of the
electric field and temperature on the hopping diffusion
coefficient. These expressions predict a violation of the traditional
form of the relation between $\mu$ and $D$ given by
Eq.~(\ref{eq-Einstein}) at very low electric fields. Since the
temperature effect on the field-dependent diffusion has been so far
left out of the scope of computer
simulations,\cite{Richert1989,Pautmeier1991} we perform here a Monte
Carlo study of the field- and temperature-dependent diffusion in the
hopping regime for 2D and 3D systems in order to check the results of
the analytical theory. Our computer simulations for the 2D and 3D
cases presented in Sec.~\ref{sec-simulation} support the analytical
theory.  Furthermore, numerical results evidence that while the
carrier mobility is stable with respect to different realizations of
disorder, the diffusion coefficient experiences significant
fluctuations from one realization to another, even for systems
containing millions of localized states. The reasons for such
different behavior between $\mu$ and $D$ is clarified and the
estimates for the system size, which is necessary to obtain stable
values of $D$, are given in Sec.~\ref{subsec-hopping-transport}.
The latter estimates show that previous numerical simulations in the
literature were performed on rather small systems, insufficient for
obtaining reliable results for the field-dependent diffusion
coefficient at low temperatures. 

In developing the analytical theory for the field-dependent
diffusion coefficient for the hopping transport regime we rely on
the analogy between the hopping transport mode and the
multiple-trapping mode.\cite{Orenstein1981,Baranovskii1997,Baranovskii2000,Rubel2004}
In Sec.~\ref{sec-theory} we present a general solution of the
problem. In Sec.~\ref{sec-comparison} the analytical
expressions from Sec~\ref{sec-theory} are compared to
simulation results. Concluding remarks are gathered in
Sec.~\ref{sec-conclusions}.

\section {Monte Carlo simulations}
\label{sec-simulation} 
In this section we study the effects of field and temperature on the
diffusion coefficient by Monte Carlo simulations in more detail than
it has been done previously, aiming at a comparison with the
analytical theory described in the following sections. Furthermore, we
study in detail the role of the system size on the simulated results
and show that in order to get reliable results for the field-dependent
diffusion at reasonably low temperatures one needs to perform
simulations on enormously large systems.

The system is modelled as a lattice of $L^2$ or $L^3$ sites with
lattice constant $d$ and the site energies $\varepsilon_i$ chosen
randomly according to Eq.~(\ref{eq-gaussian-dos}).  Periodic boundary
conditions are applied in all directions.  Hops inside a square of $7
\times 7$ sites (2D case) or a cube of $7 \times 7 \times 7$ sites (3D
case), centered at the starting site are allowed.  The simulation
proceeds as follows. A packet of $n$ non-interacting carriers is
allowed to move in the lattice until a fixed time $t$ has passed.  The
mobility $\mu$ is calculated from the average distance that the charge
carriers have moved along the direction of the field, while the
longitudinal diffusion coefficient $D_z$ is calculated from the width
of the carrier packet:
\begin{equation}
\mu = \frac {\langle z \rangle} {F t}, \qquad
D_z = \frac {\langle z^2 \rangle - \langle z \rangle ^2} {2t}.
\end{equation}
Further details of the simulation algorithm and our implementation of
it are given in Appendix \ref{sec-MC-details}.  Care was taken about
the necessary size of the simulated system in order to avoid
finite-size effects. The corresponding number of sites was $700^3$ in
the 3D case and $15000^2$ in the 2D case.

\begin{figure}
\includegraphics[width=8.6 cm]{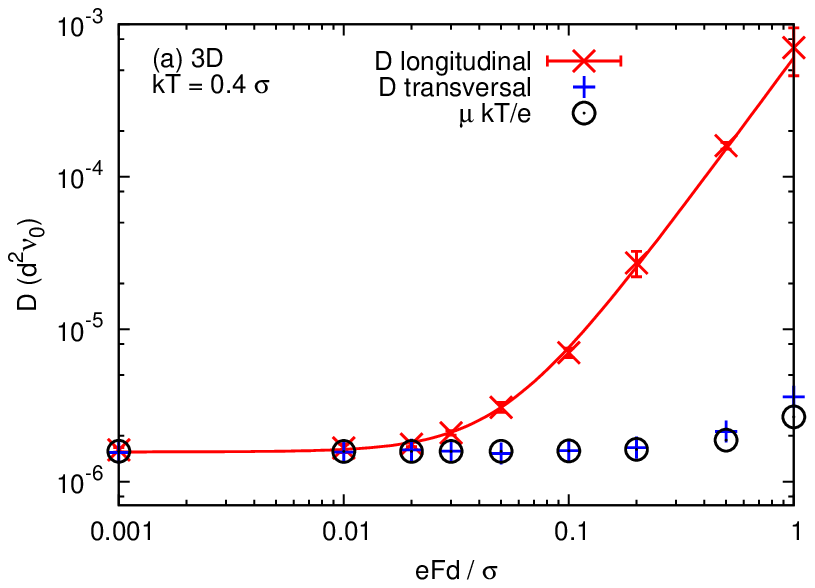}
\includegraphics[width=8.6 cm]{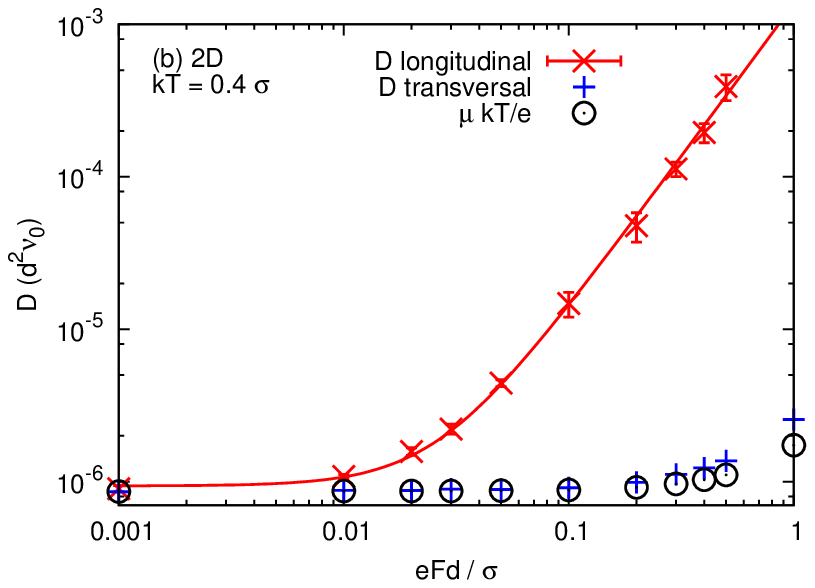}
\caption{The diffusion coefficient along and perpendicular to the
  field, and the mobility $\mu$ scaled with $kT/e$ as a function of
  the field strength, (a) in 3D, and (b) in 2D.  At small fields all
  three quantities are equal, implying that the Einstein relation is
  valid.}
\label{fig-D-and-mu}
\end{figure}

Simulation results for the diffusion coefficient both along and
perpendicular to the electric field are shown in
Fig.~\ref{fig-D-and-mu}, together with the mobility (scaled with
$kT/e$).  The localization length was $a = 0.2d$. A packet consisting
of $1000$ charge carriers was simulated for each data point, and the
simulation results were averaged over five different realizations of
disorder.  At extremely low fields it is seen that all three of the
plotted quantities are equal, which means that Einstein's relation,
Eq.~(\ref{eq-Einstein}), is valid.  With rising magnitude of the
electric field the longitudinal diffusion coefficient increases
drastically, while the mobility and transversal diffusion coefficient
remain field-independent up to much higher fields. The solid line
shows a fit to the longitudinal diffusion coefficient by a square
trial function
\begin{equation}
\label{eq-quadratic-D}
D_z(F, T) = A(T) F^2 + D_0(T).
\end{equation}
It is seen that the square function well fits the data in agreement
with the results of previous simulations.\cite{Pautmeier1991}

\begin{figure}
\includegraphics[width=8.6 cm]{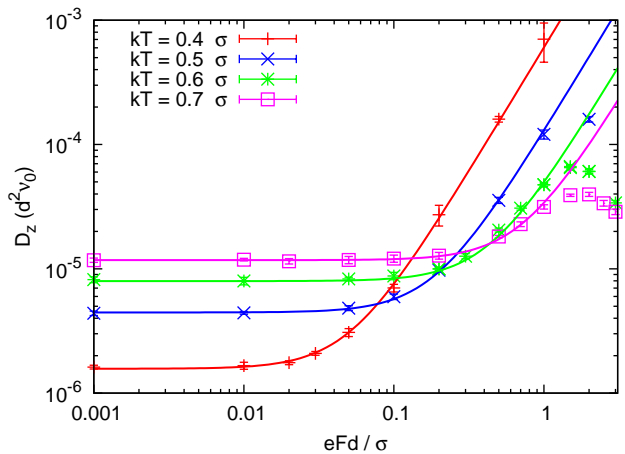}
\caption{The longitudinal diffusion coefficient $Dz$ for hopping in
  3D, as a function of the applied electric field $F$.  The solid
  lines show the best fit to the square function
  (\ref{eq-quadratic-D}).}
\label{fig-3D-Fdep}
\end{figure}
In order to study the effect of temperature on the field-dependent
diffusion coefficient, the simulations were repeated for different
temperatures. The results for the longitudinal diffusion coefficient
in 3D are collected in Fig.~\ref{fig-3D-Fdep}. The data are fitted
with the square trial functions (\ref{eq-quadratic-D}), shown in the
figure by solid lines.  The fit is good for all temperatures. The
decrease of $D_z(F)$ at very high fields for the two highest
temperatures is due to the trivial saturation effect well-known for
the one-dimensional random-energy model.\cite{Nenashev2009I} We focus
in the following on the field dependence of $D_z$ at field magnitudes
lower than the one at which $D_z$ starts to decrease with increasing
field.

\begin{figure}
\includegraphics[width=8.6 cm]{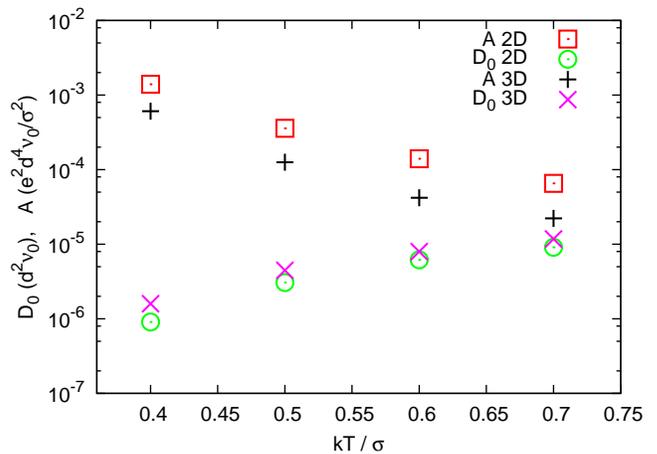}
\caption{ The temperature dependence of $D_0$, the diffusion
  coefficient at $F=0$ and $A$, obtained from Fig.~\ref{fig-3D-Fdep} for the
  localization length $a = 0.2d$.}
  \label{fig-3D-Tdep}
\end{figure}
The temperature dependences of the coefficient $A$ in the
quadratic field-dependent term for the longitudinal diffusion
coefficient $D_z$, that were obtained in the 2D and 3D cases by
fitting the simulation results in Fig.~\ref{fig-3D-Fdep} by
Eq.~(\ref{eq-quadratic-D}) are shown in Fig.~\ref{fig-3D-Tdep}
along with the field-independent term $D_0$.
Analytical expressions for $A$ will be derived in Sec.~\ref{sec-theory}.

When performing computer simulations for the field-dependent diffusion
one should be cautious with the choice of simulation parameters. The
field-induced spatial spreading of the carrier packet is caused by
trapping of some carriers onto localized states deep in energy,
while the other carriers continue their motion in shallow states being
biased by the electric field. In order to obtain reliable results for
the field-dependent diffusion coefficient, one should guarantee the
presence of such deep-in-energy states in the simulated system. Since
the DOS given by Eq.~(\ref{eq-gaussian-dos}) rapidly decreases for the
deep-in-energy states and hence the sites with deep energies are rare,
one has to simulate large systems.  The simulation time $t$
must also be chosen so large that the charge carriers have time to
visit the deep traps, which control the diffusion.
In our simulations $t$ was chosen for each temperature so that
the typical number of hops for each carrier was at least $5\cdot 10^7$.

\begin{figure}
\includegraphics[width=8.6 cm]{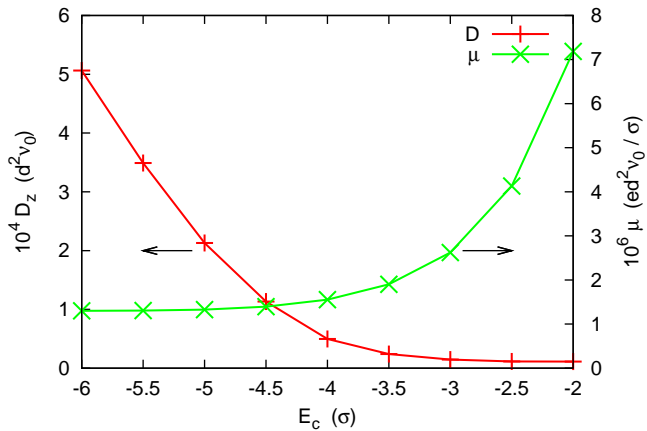}
\caption{
Diffusion coefficient $D_z$ (at the field $eFd = 0.5 \sigma$)
and mobility $\mu$ (at low field)
in a 3D system where sites with energies below $E_c$ are absent.
The temperature is $kT = 0.33 \sigma$.
}
\label{fig-cutDOS}
\end{figure}

The importance of the deep-in-energy states for the field-dependent
diffusion is demonstrated in Fig.~\ref{fig-cutDOS}, where the
values of $D_z$ (at the field $eFd = 0.5 \sigma$) and of $\mu$ (at
low fields) for the temperature $kT=0.33 \sigma$ are given when
calculated with a cut-off of the DOS below some energy $E_c$. For
these calculations the normalization of the DOS was kept, while
the states with energies below $E_c$ were excluded from the
simulation. The mobility is almost unaffected by the cutting as
long as $E_c \lesssim -4 \sigma$, while the diffusion coefficient
drastically decreases when sites with much smaller energies,
around $-6\sigma$, are removed.  The result for the mobility $\mu$
is not surprising. It has been predicted in the analytical
theory\cite{Baranovskii2000,Rubel2004} that the hopping mobility in
the Gaussian DOS for the diluted set of carriers is determined by
sites with energies in the close vicinity of the average carrier
energy $\varepsilon_\text{av} = -\sigma^2/kT$. For $kT =
0.33\sigma$, this energy is $\varepsilon_\text{av} \approx
-3\sigma$, which explains the data for $\mu$ in
Fig.~\ref{fig-cutDOS}. The result for the diffusion coefficient in
Fig.~\ref{fig-cutDOS} shows however that rare sites with even
lower energies than $\varepsilon_\text{av}$ cause the strong
dependence of the diffusion coefficient on the electric field.  In
Sec.~\ref{subsec-hopping-transport} we show that the most
important sites for the field-dependent diffusion have energies
around $\varepsilon^* = -2\sigma^2 / kT$, which is much deeper
than $\varepsilon_\text{av}$.  
This result also agrees with the data shown in
Fig.~\ref{fig-cutDOS}: for $kT = 0.33\sigma$, $\varepsilon^*
\approx -6 \sigma$. The diffusion coefficient changes drastically
when the few sites with energies below $-5.5 \sigma$ are removed,
while the mobility starts changing only when $E_c$ is in the
vicinity of $\varepsilon_\text{av}$.

In our simulations a lattice of $700^3$ sites has been used.
There are typically no sites with energies below $-6 \sigma$ in
such a system. Therefore, the $D_z(F)$ data for $kT = 0.33 \sigma$
cannot be considered as reliable in the whole simulated range of
electric fields. However, for $kT = 0.4 \sigma$, the lowest
temperature considered, $\varepsilon^* = -5 \sigma$. Energies
around $-5 \sigma$ are present in our lattice, and thus results
for $kT \gtrsim 0.4 \sigma$ can be considered as reliable for the
lattice of $700^3$ sites. It seems extremely lucky that in
previous simulations with the lattice of just $70^3$
sites\cite{Richert1989,Pautmeier1991} meaningful results were
claimed for $kT = 0.33 \sigma$.

The analytical results in the preceding paper\cite{Nenashev2009I}
were obtained for  nearest-neighbor hopping, while the
simulations above allowed also longer hops. To exclude this
difference between the models as the cause of the different field
dependencies, the simulations for two- and three-dimensional
systems were repeated with only nearest-neighbor hopping allowed.
The mobility and diffusion coefficient obtained in this case were
somewhat lower than those for the variable range hopping, however
the parabolic shape of the field dependence for the diffusion
coefficient did not change.

Our numerical results can be qualitatively summarized as follows:
(i) the diffusion coefficient \emph{along} the electric field depends
  parabolically on the field strength $F$ at low fields;
(ii) the diffusion coefficient \emph{perpendicular} to the field is
  field-independent in the range of fields where Ohm's law is
  fulfilled;
(iii) the field-dependent part of the diffusion coefficient rapidly
  decreases with increasing temperature;
(iv) the field-dependent part of the diffusion coefficient is very
  sensitive to sites which energies are lower than the mean carrier
  energy.

The discussion above, in particular the latter statement on the
decisive role of rare sites with very deep energies that can
hardly be found in the finite simulation arrays, raises the task of
developing an analytical theory for the diffusion process in the
hopping regime enhanced by an electric field. There is no such
theory in the literature so far. The only relevant theory is the
one developed by Rudenko and Arkhipov for \emph{band
transport} in materials with traps.\cite{Rudenko1982} Although
this theory predicts a parabolic field dependence of the
longitudinal diffusion coefficient, it cannot be directly applied
to \emph{hopping transport}, because it operates with quantities
that are specific for band conductivity (for example, the effective
density of states in the conduction band). Therefore it is
necessary to develop a theory for hopping transport, particularly
because Monte Carlo simulations suffer from finite-size
effects as described above.  We will give such a theory in
Sec.~\ref{sec-theory}.

\section{Analytical results}
\label {sec-theory}

\subsection{Approximation of independent jumps \\(general consideration)}
\label{subsec-independent-general}

\begin{figure}
\includegraphics[width=8.6 cm]{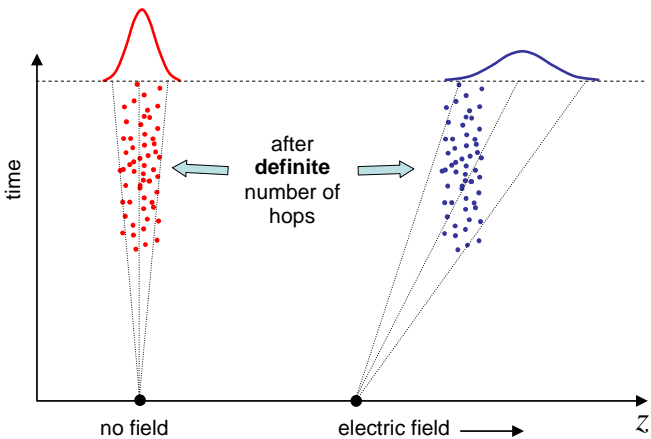}
\caption{A sketch of spatio-temporal distribution of carriers after
  some \emph{definite} number of jumps.}
\label{fig-sketch}
\end{figure}

The origin of the field-dependent diffusion can be understood
qualitatively with the aid of a spatio-temporal picture of the
carrier distribution sketched in Fig.~\ref{fig-sketch}. The small
dots indicate the scatter of carriers \emph{after some definite
number of jumps}, assuming that all carriers start at the same
time $t=0$ from the same point. To get the spatial distribution of
carriers \emph{at some definite time} $t^*$, one may ``project''
these dots from the starting point to the line $t=t^*$. The
direction of ``projecting'' is determined by the drift velocity.
When the drift velocity is equal to zero (no electric field, the
left part of Fig.~\ref{fig-sketch}), the spatial distribution of
carriers at $t=t^*$ does not depend on the scatter of times spent
by the carriers in order to perform a fixed number of jumps.  On
the contrary, in the case of drift caused by an electric field (the
right part of Fig.~\ref{fig-sketch}), this scatter in times
``projects'' into the line $t=t^*$, which gives rise to the
broadening of the spatial distribution at $t=t^*$. This broadening
is the reason for the enhancement of the diffusion coefficient due
to electric field.

The above consideration shows that fluctuations of the durations
of jumps are responsible for the field-induced diffusion. These
fluctuations can be especially pronounced for systems with a broad
distribution of site energies, because jumps from energetically
deep sites to transport sites demand exponentially long times.

Let us start from a general form of the \emph{quantitative}
description for the field dependence of the diffusion coefficient.
Our consideration is based on the assumption that \emph{successive
jumps are statistically independent}, i.~e.\ the diffusion process
is Markovian. The latter means that increments in carrier
coordinates at a given jump, as well as the time interval from the
preceding jump till the given one, do not depend on the carrier
prehistory. This assumption definitely does not hold for the
one-dimensional hopping transport considered in the previous
paper.\cite{Nenashev2009I} In the 1D case, the probability of returning to
an already visited trap is not negligible, and hence
the consequent jumps must be correlated. Therefore our analytical
consideration based on the assumption of the statistical
independence of the successive jumps present below can be valid
only for 2D and 3D cases since it is reasonable to assume that in
the latter cases the carrier trajectories are non-returning.

Let $t^*$ be some fixed moment of time ($t^*$ is large compared to
the mean time of a jump); $x^*$, $y^*$ and $z^*$ be the carrier's
displacements along the axes $X$, $Y$ and $Z$ during the time
interval $[0,t^*]$; $x_1$, $y_1$, $z_1$ and $t_1$ be coordinate
displacements and a time increment related to only one jump,
respectively; $x_n$, $y_n$, $z_n$ be the displacements after $n$
successive jumps. Angle brackets will denote averaging over jumps
performed by different carriers,  equivalent to the averaging over
successive jumps of one carrier.

We start with the simple case of zero electric field. Since there
is no drift, the expectation values of $x^*$, $y^*$ and $z^*$
vanish, and one obtains for the diffusion coefficients $D_x$,
$D_y$, $D_z$
\begin{equation}
  D_x = \lim_{t^*\rightarrow\infty} \frac{\langle x^{*2}\rangle}{2t^*}, \,
  D_y = \lim_{t^*\rightarrow\infty} \frac{\langle y^{*2}\rangle}{2t^*}, \,
  D_z = \lim_{t^*\rightarrow\infty} \frac{\langle z^{*2}\rangle}{2t^*}.
\end{equation}
For large $t^*$, the displacements $x^*$, $y^*$, $z^*$ are
approximately equal to $x_N$, $y_N$, $z_N$ -- displacements after $N$
jumps, where $N=t^*/\langle t_1\rangle$ is the mean number of jumps
during the time $t^*$. Consequently,
\begin{equation}
\frac{\langle x^{*2}\rangle}{2t^*} \approx \frac{\langle
  x_N^2\rangle}{2t^*} = N\frac{\langle x_1^2\rangle}{2t^*} =
\frac{\langle x_1^2\rangle}{2\langle t_1\rangle}.
\end{equation}
In the limit $t^*\rightarrow\infty$ this approximate equality becomes
exact, and one gets
\begin{equation} \label{eq-D_zero_field}
  D_x = \frac{\langle x_1^2\rangle}{2\langle t_1\rangle}, \quad
  D_y = \frac{\langle y_1^2\rangle}{2\langle t_1\rangle}, \quad
  D_z = \frac{\langle z_1^2\rangle}{2\langle t_1\rangle}.
\end{equation}

Let us now consider the case of a finite electric field along the
$Z$-axis. Since the expectation values of $x^*$ and $y^*$ are
still zero, all the considerations above remain valid, and
Eq.~(\ref{eq-D_zero_field}) remains correct with respect to $D_x$
and $D_y$. One cannot obtain $D_z$ by literally the same
way, because $\langle z^*\rangle\neq0$. Instead, one may apply
this argumentation to a variable $\tilde{z}=z-vt$, where
$v=\langle z_1\rangle/\langle t_1\rangle$ is the drift velocity.
Since $\langle\tilde{z}^*\rangle = \langle\tilde{z}_N\rangle =
\langle\tilde{z}_1\rangle = 0$, one gets
\begin{equation}
D_z \equiv D_{\tilde z} \! = \!\! \lim_{t^*\rightarrow\infty} \!\!
\frac{\langle \tilde z^{*2}\rangle}{2t^*} \! = \!  \frac{\langle
  \tilde z_1^2\rangle}{2\langle t_1\rangle} = \frac{\langle
  z_1^2\rangle \! - \! 2v\langle z_1t_1\rangle \! + \! v^2\langle
  t_1^2\rangle}{2\langle t_1\rangle},
\end{equation}
and finally
\begin{equation} \label{eq-D_nonzero_field}
  D_z = \frac{\langle
    z_1^2\rangle}{2\langle t_1\rangle} - \frac{\langle
    z_1\rangle\langle z_1t_1\rangle}{\langle t_1\rangle^2} +
  \frac{\langle z_1\rangle^2\langle t_1^2\rangle}{2\langle
    t_1\rangle^3}.
\end{equation}

Since the mean values $\langle z_1\rangle$, $\langle z_1^2\rangle$
and $\langle z_1t_1\rangle$ depend on the electric field,
Eq.~(\ref{eq-D_nonzero_field}) describes the field-dependent
diffusion along the field direction. On the other hand, there is
no reason for the mean values $\langle x_1^2\rangle$, $\langle
y_1^2\rangle$ and $\langle t_1\rangle$ to be field-dependent in
small electric fields. Therefore, according to
Eq.~(\ref{eq-D_zero_field}), the transversal diffusion
coefficients $D_x$ and $D_y$ are expected to be constant inside
the Ohmic regime.

Let us discuss the shape of the dependence $D_z(F)$ near the point
$F=0$. Since $\langle z_1\rangle$ is proportional to the electric
field ($\langle z_1\rangle=\mu F \langle t_1\rangle$, where $\mu$ is
mobility), the \emph{third} term in the r.-h. side of
Eq.~(\ref{eq-D_nonzero_field}) gives a contribution to $D_z(F)$ that
is quadratic in $F$. In the \emph{first} term,
\begin{equation}
  \langle z_1^2\rangle = \langle z_1\rangle^2 + \sigma^2[z_1],
\end{equation}
where $\sigma[a]$ denotes the standard deviation of a random
variable $a$. Since $\sigma[z_1]$ is not sensitive to the electric
field, the first term in the r.-h. side of
Eq.~(\ref{eq-D_nonzero_field}) is a sum of a constant and a term
quadratic in $F$. The behavior of the \emph{second} term in
the r.-h. side of Eq.~(\ref{eq-D_nonzero_field}) depends on the
symmetry of the system. If in the absence of electric field
the directions $Z$ and $-Z$ are equivalent, then $\langle
z_1t_1\rangle=0$ at $F=0$. In this case one expects that $\langle
z_1t_1\rangle\sim F$ and consequently the second term in the r.-h.
side of Eq.~(\ref{eq-D_nonzero_field}) is quadratic in $F$.
However if the positive and negative directions along the $Z$-axis
are non-equivalent, $\langle z_1t_1\rangle$ can be non-zero at
$F=0$, which gives a contribution to the diffusion coefficient
linear in $F$.

In summary, if the directions $Z$ and $-Z$ are equivalent, which we
will assume in the following,  the longitudinal diffusion
coefficient $D_z$ at small fields is described by
Eq.~(\ref{eq-quadratic-D}) where $D_0$ and $A$ are
field-independent coefficients.

\subsection{Multiple-trapping band conductivity}
\begin{figure}
\includegraphics[width=8.6 cm]{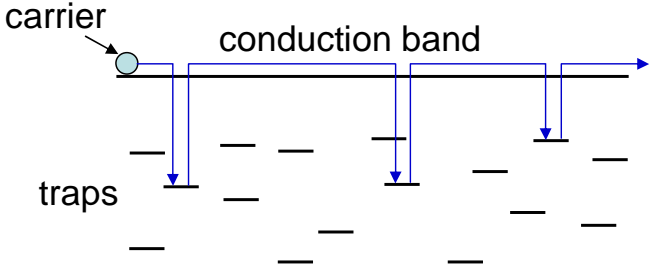}
\caption{Multiple-trapping conductivity.}
\label{fig-multi-trapping}
\end{figure}

Let us first apply our approach to the multiple-trapping (MT)
model of \emph{band} conductivity.  The model includes processes
of capture of free carriers by traps, emission of trapped
carriers, and free motion (Brownian motion plus drift in the electric
field) of carriers in the band (Fig.~\ref{fig-multi-trapping}).
The band is characterized by the effective density of states
$N_c$, the mobility $\mu_f$ and the diffusion coefficient $D_f$ of
free carriers. The traps are characterized by the density of
states $g(\varepsilon)$ and the capture rate (for unit free
carrier concentration) $c(\varepsilon)$. The energy $\varepsilon$
is counted from the band edge.

The problem of the field-effect on the diffusion coefficient in
the MT model has been considered by Rudenko and
Arkhipov,\cite{Rudenko1982} who treated the evolution in time of
the one-dimensional carrier distribution function. In
Ref.~\onlinecite{Rudenko1982}, transport was assumed to take place in
quasi-equilibrium, i.~e., the distribution function was assumed to
change slowly in comparison with the rate of exchange between
traps and the conduction band. Our approach is free of the
assumption of quasi-equilibrium transport. Besides, our
consideration provides information not only about the longitudinal
diffusion coefficient, but also about the transversal one.

Instead of examining the carrier distribution function, we will
follow the motion of a \emph{single} carrier and consider the
statistics of this motion. In order to use the expressions
(\ref{eq-D_zero_field}) and (\ref{eq-D_nonzero_field}), we represent
the motion of a carrier as a sequence of ``jumps'', each of them
beginning at the moment of escape from a trap. Successive ``jumps''
are statistically independent, because the processes of capture
take place in different points in space and there is no
correlation between them.

Each ``jump'' of a carrier consists of two contributions: free
motion and sitting on a trap. Let $t_{1f}$ be the time of free
motion (between emission and capture), and $t_{1t}$ be the time of
being trapped (between capture and emission); $t_{1f}$ and
$t_{1t}$ are independent random variables. We denote their
expectation values $\langle t_{1f}\rangle$ and $\langle
t_{1t}\rangle$ as $T_1$ and $T_2$, respectively. Since $T_1$ is
the lifetime of free carriers, its reciprocal value is the
capture rate:
\begin{equation}
  T_1^{-1} = \int c(\varepsilon)g(\varepsilon)d\varepsilon.
\end{equation}
The variable $t_{1f}$ obeys an exponential distribution. Hence its
mean square $\langle t_{1f}^2\rangle$ is equal to $2T_1^2$.

Collecting all this information, one gets the following
statistical expressions for a single ``jump'' (electric field $F$
is directed along $Z$):
\begin{align}
   \langle t_1 \rangle &= T_1+T_2, \notag \\
   \langle t_1^2 \rangle &= \langle t_{1f}^2 \rangle + 
   2\langle t_{1f}t_{1t} \rangle + \langle t_{1t}^2 \rangle = \!
  2T_1(T_1+T_2)+\langle t_{1t}^2 \rangle, \notag \\
   \langle x_1 \rangle &= \langle y_1 \rangle = 0, \\
   \langle z_1 \rangle &= \mu_fF \langle t_{1f} \rangle = \mu_fFT_1, \notag \\
   \langle x_1^2 \rangle &= \langle y_1^2
  \rangle = 2D_f \langle t_{1f} \rangle= 2D_fT_1,  \notag \\
   \langle z_1^2 \rangle &= \mu_f^2F^2 \langle t_{1f}^2 \rangle
  + 2D_f \langle t_{1f} \rangle= 2\mu_f^2F^2T_1^2+2D_fT_1, \notag \\
   \langle z_1t_1 \rangle &= \langle z_1t_{1f} \rangle +
  \langle z_1t_{1t} \rangle = 2\mu_fFT_1^2+\mu_fFT_1T_2. \notag
\end{align}
Using these mean values, one can evaluate the mobility:
\begin{equation} \label{eq-RA-mu1} \mu = \frac{\langle z_1 \rangle}{F
    \langle t_1 \rangle} = \mu_f\frac{T_1}{T_1+T_2},
\end{equation}
the transversal diffusion coefficient via Eq.~(\ref{eq-D_zero_field})
\begin{equation} \label{eq-RA-Dxy1} D_x = D_y = \frac{\langle x_1^2
    \rangle}{2 \langle t_1 \rangle} = D_f\frac{T_1}{T_1+T_2},
\end{equation}
and the longitudinal diffusion coefficient via Eq.~(\ref{eq-D_nonzero_field})
\begin{equation} \label{eq-RA-Dz1} D_z = D_f\frac{T_1}{T_1+T_2} +
  \mu_f^2F^2\frac{T_1^2 \langle t_{1t}^2 \rangle}{2(T_1+T_2)^3}.
\end{equation}
The parameters $T_1$, $T_2$, and $\langle{t}_{1t}^2\rangle$ are
governed by the capture cross-sections and emission rates of the
traps. For moderate electric fields, these cross-sections and
rates can be regarded as field-independent, and consequently one
can use \emph{equilibrium} values for $T_1$, $T_2$, and
$\langle{t}_{1t}^2\rangle$. This gives an opportunity to simplify
Eqs.~(\ref{eq-RA-mu1})--(\ref{eq-RA-Dz1}). The
ratio $\alpha\equiv T_1/(T_1+T_2)$ contributing to these equations
is simply the fraction of free carriers in the equilibrium state.
The distribution of the dwell times $\tau$ of a carrier at
some individual trap is exponential with the mean value
$\langle\tau\rangle=1/\Gamma_\uparrow$, where $\Gamma_\uparrow$ is
the emission rate. Consequently, the mean square of this dwell
time is $\langle\tau^2\rangle=2\Gamma_\uparrow^{-2}$. The mean
value $\langle{t}_{1t}^2\rangle$ is a weighted average of values
$\langle\tau^2\rangle$, where the probability of visiting a trap
serves as the weight. Therefore,
\begin{equation}
\langle{t}_{1t}^2\rangle = \int 2\Gamma_\uparrow^{-2}(\varepsilon)
p(\varepsilon) d\varepsilon,
\end{equation}
where $p(\varepsilon)d\varepsilon$ is the probability of visiting (at a
given ``jump'') a trap with an energy in the range
$[\varepsilon;\varepsilon+d\varepsilon]$:
\begin{equation}
p(\varepsilon)d\varepsilon =
\frac{c(\varepsilon)g(\varepsilon)d\varepsilon}{\int
  c(\varepsilon)g(\varepsilon)d\varepsilon} \equiv T_1
c(\varepsilon)g(\varepsilon)d\varepsilon.
\end{equation}
Expressing the escape rates through capture cross-sections:
\begin{equation}
  \Gamma_\uparrow(\varepsilon) = c(\varepsilon) N_c e^{\varepsilon/kT},
\end{equation}
one obtains $\langle{t}_{1t}^2\rangle = 2T_1I$, where
\begin{equation} \label{eq-RA-I} I = \int
  \left(\frac{e^{-\varepsilon/kT}}{c(\varepsilon)N_c}\right)^2
  c(\varepsilon)g(\varepsilon)d\varepsilon.
\end{equation}
Finally, Eqs.~(\ref{eq-RA-mu1})--(\ref{eq-RA-Dz1}) get the
following form:
\begin{equation} \label{eq-RA-final}
  \mu = \alpha\mu_f, \quad
  D_{xy}=\alpha D_f, \quad
  D_z=\alpha D_f + \alpha^3I\mu_f^2F^2,
\end{equation}
where $\alpha$ is the equilibrium fraction of free carriers, and $I$ is
the integral defined by Eq.~(\ref{eq-RA-I}).

The expressions for $\mu$ and $D_z$ are the same as the ones obtained by
Rudenko and Arkhipov.\cite{Rudenko1982} However, our consideration
provides some new information about diffusion in the multiple-trapping
conductivity regime. First, Eq.~(\ref{eq-RA-final}) shows that the
\emph{transversal} diffusion coefficient is field-independent. Second,
we have proven that the equations for $\mu$ and $D_z$ are \emph{exact} 
(provided that emission/capture probabilities are not influenced by
electric field), whereas in Ref.~\onlinecite{Rudenko1982} they were
obtained only under so-called ``quasi-equilibrium conditions''. 

\subsection{Hopping transport}
\label{subsec-hopping-transport} Now we will apply
Eq.~(\ref{eq-D_nonzero_field}) to two- and three-dimensional
hopping transport in a system with a Gaussian DOS, in a manner
very similar to our consideration of the multiple-trapping
conductivity.

Our analysis is based on the following hypothesis: \emph{the field
  dependence of the diffusion coefficient in 2D and 3D hopping is mainly due to
  very rare and energetically deep sites.} We will call them
\emph{``traps''.} The traps are rare in two senses: first, the
typical distance between the traps is large in comparison to the
inhomogeneities of the mobility; second, the probability of being
trapped is small, so that the traps do not affect the carrier mobility.
``Energetically deep'' means that the energies of the traps are
far below the mean energy of the carriers.
We will show below that this hypothesis provides a reasonable
description of the Monte Carlo simulation results on the
field-dependent diffusion.

As in the previous subsection, we consider the motion of a carrier as
a sequence of ``jumps'', each beginning when the carrier enters a
trap, and ending when the carrier enters \emph{another} trap.  The
durations of different ``jumps'' are not correlated. The same is true
for the carrier displacements. Indeed, the three-dimensional Brownian
motion is non-returning. This property guarantees that the
carrier always visits a \emph{new} trap, and that the trajectories of its
motion between different traps do not overlap. Hence there are no
reasons for correlations between successive ``jumps''. This is the
point where dimensionality is important. In one dimension, 
the probability of visiting a previously visited trap is
not negligible. Consequently the ``jumps'' must be correlated. Since the
``jumps'' in the 3D case are not correlated, one can obtain the
mobility $\mu$ and the diffusion coefficient $D$ from the statistics
of ``jumps'' using the method of
Sec.~\ref{subsec-independent-general}.  For the 2D case we also
use the assumption of independent ``jumps''.

As in the case of multiple-trapping conductivity, each ``jump''
consists of two contributions: ``free'' motion of a carrier and
``sitting`` on a trap. Let $t_{1f}$ and $t_{1t}$ be the durations
of these contributions; $T_1$ and $T_2$ denoting the mean values
$\langle t_{1f}\rangle$ and $\langle t_{1t}\rangle$, respectively
(where averaging is over successive ``jumps''); $\mu_f$ and $D_f$
be the mobility and the diffusion coefficient of ``free'' motion
(energetically far above the traps). With these notations one can
follow the same derivation as in the previous subsection and see
that Eqs.~(\ref{eq-RA-mu1})--(\ref{eq-RA-Dz1}) are still valid
also for hopping transport. It is convenient to rewrite these
equations:
\begin{equation} \label{eq-3DG-parabolic}
 \mu=\mu_0, \quad D_x=D_y=D_0, \quad D_z=D_0+AF^2,
\end{equation}
where
\[
 \mu_0 = \mu_f \frac{T_1}{T_1+T_2}, \qquad D_0 = D_f \frac{T_1}{T_1+T_2},
\]
\begin{equation} \label{eq-3DG-A1}
 A=\mu_0^2\frac{\langle t_{1t}^2 \rangle}{2(T_1+T_2)}.
\end{equation}
In the regime of ohmic conductivity, the values of $\mu_f$, $D_f$,
$T_1$, $T_2$ and $\langle t_{1t}^2 \rangle$ can be regarded as
field-independent 
because a sufficiently small electric field does not significantly
perturb the probabilities of capture and release. Therefore one
can neglect a possible dependence of $\mu_0$, $D_0$ and $A$ on the
electric field. In the following we will use the zero-field values
for these quantities.

Let us calculate the coefficient $A$. For convenience, we
will consider the system as a large but finite one (with periodic
boundary conditions, to allow drift). Then one can obtain the
following expressions for $T_1$, $T_2$ and $\langle t_{1t}^2
\rangle$:
\begin{equation} \label{eq-3DG-T1}
 T_1 = \left( \sum\limits_t p_t \Gamma_{\text{esc},t} \right)^{-1},
\end{equation}
\begin{equation} \label{eq-3DG-T2}
 T_2 = T_1 \sum\limits_t p_t \,,
\end{equation}
\begin{equation} \label{eq-3DG-t1f2}
 \langle t_{1t}^2 \rangle = 2 T_1 \sum\limits_t p_t \Gamma_{\text{esc},t}^{-1} \,,
\end{equation}
where the index $t$ runs over all traps, $p_t$ is the probability that a
carrier is at site $t$, and $\Gamma_{\text{esc},t}$ is the rate of escaping
from the trap $t$.

Eq.~(\ref{eq-3DG-T1}) results from a consideration of the carrier flow
from/to traps. Namely, the flow of carriers out of traps is equal to
$\sum_t p_t \Gamma_{\text{esc},t}$; the flow into traps is $T_1^{-1}$. In
equilibrium, these flows are equal to each other, which gives
Eq.~(\ref{eq-3DG-T1}).

In order to obtain Eqs.~(\ref{eq-3DG-T2}), (\ref{eq-3DG-t1f2}), we
introduce the probability $\mathcal P_t$ that trap $t$ will be the
next visited trap. Let us consider the balance of flows from/to
trap $t$. Flow from this trap is equal to $p_t \Gamma_{\text{esc},t}$.
Flow to it is $\mathcal P_t T_1^{-1}$. Therefore
\begin{equation} \label{eq-3DG-Pt}
 \mathcal P_t = T_1 p_t \Gamma_{\text{esc},t}.
\end{equation}
The mean time of being captured at trap $t$ is $\Gamma_{\text{esc},t}^{-1}$. To
obtain $T_2$, we average these mean times with corresponding weights
$\mathcal P_t$:
\begin{equation}
 T_2 = \sum\limits_t \Gamma_{\text{esc},t}^{-1} \mathcal P_t \,.
\end{equation}
Substituting here Eq.~(\ref{eq-3DG-Pt}), one obtains
Eq.~(\ref{eq-3DG-T2}). Analogously, the mean \emph{square} of the time
of being trapped at site $t$ is $2\Gamma_{\text{esc},t}^{-2}$ (the factor of
2 comes from the exponential distribution of dwell times). Again, we
take a weighted average to obtain $\langle t_{1t}^2 \rangle$:
\begin{equation}
 \langle t_{1t}^2 \rangle = \sum\limits_t 2\Gamma_{\text{esc},t}^{-2} \mathcal P_t \,,
\end{equation}
that, together with Eq.~(\ref{eq-3DG-Pt}), provides Eq.~(\ref{eq-3DG-t1f2}).

The substitution of Eqs.~(\ref{eq-3DG-T1})--(\ref{eq-3DG-t1f2}) into
Eq.~(\ref{eq-3DG-A1}) gives the following expression for $A$:
\begin{equation} \label{eq-3DG-A1.5}
  A = \mu_0^2 \,
  \frac{\sum\limits_t p_t \Gamma_{\text{esc},t}^{-1}}{1+\sum\limits_t p_t}
  \,.
\end{equation}
Let us try to simplify this expression. The sum in the denominator
is the probability that a carrier is at some trap. Due to the
rarity of traps, this probability is small, and it can be
neglected. The summation in the numerator will be taken over
\emph{all} sites:
\begin{equation}
 A \approx \mu_0^2 \sum\limits_s p_s \Gamma_{\text{esc},s}^{-1} \,.
\end{equation}
 We will use curly braces to denote the summation in which index $s$ runs over
all sites:
\begin{equation}
 \{a\} \equiv \sum\limits_s p_s a_s,
\end{equation}
where $a$ is any quantity specific for sites. It is obvious that
curly braces mean averaging over an ensemble of particles (or time
averaging, which is the same for finite system and long enough
time). Thus,
\begin{equation}
 A \approx \mu_0^2 \, \{ \Gamma_{\text{esc}}^{-1} \} \,.
\end{equation}

Then, the rate of escape $\Gamma_{\text{esc},s}$ is related to the sum
$\sum_{s'\neq s} \Gamma_{ss'}$ of transition rates from site $s$
to any other sites:
\begin{equation}
 \Gamma_{\text{esc},s} = n_{\text{esc},s}^{-1} \sum\limits_{s'\neq s} \Gamma_{ss'} \,,
\end{equation}
where $n_{\text{esc},s}$ is the mean number of ``attempts'' to escape
from site $s$ (including the successful one). These ``attempts''
are events of carrier's hopping out of site $s$ before the carrier
moves so far from this site that it completely forgets the
prehistory related to this site. Let us denote the sum
$\sum_{s'\neq s} \Gamma_{ss'}$ as $t_s^{-1}$. Then one can rewrite
$A$ as
\begin{equation}
 A \approx \mu_0^2 \, \{ n_{\text{esc}} t \} \,,
\end{equation}
or, introducing the averaged number of escape attempts $\overline{n_{\text{esc}}}$,
\begin{equation} \label{eq-3DG-A2}
 A \approx \overline{n_{\text{esc}}} \, \mu_0^2 \, \{ t \} \,.
\end{equation}

Calculating $\{t\}$ implies averaging over site energies
($\varepsilon_s$) and over quantities related to the neighborhood
(energies of neighbors $\varepsilon_{s'}$ and distances to them
$r_{ss'}$). It is possible to treat these two kinds of averaging
\emph{separately}, using our assumption that ``optimal'' traps are
deep in energy. Since the traps are deep, each hop from a trap is
upward in energy. Therefore the Miller-Abrahams rates,
Eq.~(\ref{eq-Miller-Abrahams}), for such hops can be written as
\begin{equation}
\Gamma_{ss'} = \nu_0 \exp\left( -\frac{2r_{ss'}}{a}
  +\frac{\varepsilon_s-\varepsilon_{s'}}{kT} \right) .
\end{equation}
The dependence of this expression on $\varepsilon_s$ has the form
of a factor $\exp(\varepsilon_s/kT)$. Hence, the quantity $t_s =
(\sum_{s'} \Gamma_{ss'})^{-1}$ can be factorized (only for deep
sites $s$) as
\begin{equation}
 t_s = \tau_s \, \exp(-\varepsilon_s/kT),
\end{equation}
where $\tau_s$ does not depend on $\varepsilon_s$:
\begin{equation} \label{eq-3DG-tau-s}
  \tau_s = \nu_0^{-1} \, \left[
    \sum\limits_{s'\neq s} \exp\left( -\frac{2r_{ss'}}{a}
      -\frac{\varepsilon_{s'}}{kT} \right) \right]^{-1}.
\end{equation}
Since there are no correlations between $\exp(-\varepsilon_s/kT)$
and $\tau_s$, one can average these quantities separately:
\begin{equation}
\{ t \} \approx \{ e^{-\varepsilon_s/kT} \tau_s \} = \{
e^{-\varepsilon_s/kT} \} \, \tau.
\end{equation}
Here $\tau$ is the mean value of $\tau_s$ (an arithmetical average
over all $s$). Now, the calculation of $\{ e^{-\varepsilon_s/kT} \}$
is simple. Remember that $p_s$ is an equilibrium probability of
finding a carrier at site $s$:
\begin{equation}
 p_s = \frac{e^{-\varepsilon_s/kT}}{\sum\limits_{s'}e^{-\varepsilon_{s'}/kT}}.
\end{equation}
Therefore
\begin{equation}
 \{ e^{-\varepsilon_s/kT} \} \equiv \sum\limits_s p_s e^{-\varepsilon_s/kT} =
\end{equation}
\begin{equation} \label{eq-3DG-averaging-integrals}
  \frac{\sum\limits_s
    e^{-2\varepsilon_s/kT}}{\sum\limits_{s'}e^{-\varepsilon_{s'}/kT}}
  = \frac{\int e^{-2\varepsilon/kT} g(\varepsilon) d\varepsilon}{\int
    e^{-\varepsilon/kT} g(\varepsilon) d\varepsilon} .
\end{equation}
For a Gaussian DOS $g(\varepsilon)$ given by
Eq.~(\ref{eq-gaussian-dos}), one obtains
\begin{equation}
 \{ e^{-\varepsilon_s/kT} \} = \exp\left( \frac{3\sigma^2}{2(kT)^2} \right) .
\end{equation}
Hence
\begin{equation} \label{eq-3DG-t-tau}
  \{ t \} \approx \tau \exp\left( \frac{3\sigma^2}{2(kT)^2} \right) .
\end{equation}
Substituting this result into Eq.~(\ref{eq-3DG-A2}), one obtains
\begin{equation} \label{eq-3DG-A3}
 A \approx \overline{n_{\text{esc}}} \, \mu_0^2\, \tau \exp\left( \frac{3\sigma^2}{2(kT)^2} \right) .
\end{equation}
This equation contains $\tau$. The proper way to calculate $\tau$
is the numerical one: to generate at random a large enough number
of neighborhoods of the given site (i.~e.\ sets of energies
$\varepsilon_{s'}$ of neighboring sites and distances $r_{ss'}$
from the given site $s$), to calculate $\tau_s$ for each
neighborhood according to Eq.~(\ref{eq-3DG-tau-s}), and then to
average the results for $\tau_s$.
In the next section we compare Eq.~(\ref{eq-3DG-A3}) to the
results of Monte Carlo simulations from 
Sec.~\ref{sec-simulation}.

Let us now reveal which traps are ``optimal'' in the sense that
they determine the diffusion coefficient. The integrand
in the numerator of Eq.~(\ref{eq-3DG-averaging-integrals}) has a
sharp peak of width $\sigma$ at $\varepsilon=-2\sigma^2/kT$. Hence the
``optimal'' traps are sites with energies in the range
$-2\sigma^2/kT\pm\sigma$.

It is now possible to justify our assumption about the possibility
to distinguish between the transport sites and the traps. We restrict
ourselves to the case of low temperatures, $kT\ll\sigma$. The
typical energy of the ``optimal'' traps, $-2\sigma^2/kT$, is
significantly lower that the mean energy of carriers
$\{\varepsilon\}=-\sigma^2/kT$ (the difference is much larger than
$kT$). Herewith the assumption that traps are deep in energy is
fulfilled. The probability for a carrier to have an energy lower
than $-2\sigma^2/kT$ is $\sim\exp(-\sigma^2/2(kT)^2)$. This value
is much smaller than unity. Therefore neglecting the sum in the
denominator of Eq.~(\ref{eq-3DG-A1.5}) is justified. Let us now
find the typical distance $L_t$ between ``optimal'' traps. Their
concentration can be estimated as the concentration of sites with
energies lower than $-2\sigma^2/kT$, which is about $\sim
N\exp(-2\sigma^2/(kT)^2)$, where $N$ is the total concentration of
sites. Hence $L_t\sim N^{-1/3}\exp(2\sigma^2/3(kT)^2)$. We should
compare it to the characteristic size $L_\mu$ of inhomogeneities
of the mobility. One can estimate this size as a typical distance
between sites that are important for the mobility. These are sites
with energies close to the mean energy of 
carriers,\cite{Baranovskii2000,Rubel2004}
$-\sigma^2/kT$. Their concentration is about
$N\exp(-\sigma^2/2(kT)^2)$. Hence $L_\mu\sim
N^{-1/3}\exp(\sigma^2/6(kT)^2)$. Herewith we obtain the strong
inequality $L_t\gg L_\mu$, which implies that it is safe to
describe the motion of the carriers between traps by means of a
constant mobility $\mu_f$ and a diffusion coefficient $D_f$.

All of the above considerations were based on the Boltzmann statistics
for the charge carriers. Let us discuss the effect of the carrier
concentration $n$ on the obtained results.
The usage of Boltzmann statistics is justified if the
Fermi level $\varepsilon_F$ is far below the energies of the sites
that make a major contribution to the mean value $\{t\}$. As
described above, the main contribution to $\{t\}$ comes from sites
with energies in the vicinity of $-2\sigma^2/kT$. A simple
calculation shows that carrier concentration $n_\text{diff}$
corresponding to a Fermi level $\varepsilon_F=-2\sigma^2/kT$ is
\begin{equation} \label{eq-3DG-ndiff}
 n_\text{diff} \approx N \exp\left( -\frac{3\sigma^2}{2(kT)^2} \right),
\end{equation}
where $N$ is the concentration of sites. The theory considered
above presumes that $n\ll n_\text{diff}$. In this limit, the
diffusion coefficient does not depend on the carrier
concentration. For larger concentrations, $n\gg n_\text{diff}$,
sites with energies $\approx-2\sigma^2/kT$ are essentially
occupied. In the latter case they cannot efficiently capture the
moving charge carriers and therefore the contribution of such
sites to the diffusion process (in particular to $\{t\}$) is
suppressed. According to Eq.~(\ref{eq-3DG-A2}), the field-induced
part of the diffusion coefficient $D_z(F)-D_0$ is proportional to
$\{t\}$. Therefore, in the case $n\gg n_\text{diff}$, $D_z(F)-D_0$
decreases with increasing $n$. Detailed consideration of this
concentration-dependent diffusion is beyond the scope of the
present paper.

We should also note that if the sample contains less than
$\sim\exp(2\sigma^2/(kT)^2)$ sites, then the average number of
``optimal'' traps in the sample is less than 1. One should
therefore expect large sample-to-sample variations of $D_z(F)-D_0$
in this case.

\section{Comparison of Monte Carlo and analytical results}
\label {sec-comparison} In this section we discuss relations
between the simulation results described in Sec.~\ref{sec-simulation}
and the analytical expression~(\ref{eq-3DG-A3}).

There are three quantities in Eq.~(\ref{eq-3DG-A3}) that are not
input parameters of the model: the zero-field mobility $\mu_0$,
the averaged number of escape attempts $\overline{n_{\text{esc}}}$, and the
value of $\tau$. For the mobility, we will use the values taken from
our simulations. There is no need for a special discussion of the
mobility, because it is a well-studied property of the transport
model considered here.\cite{Baranovski2006,Baranovskii2000,Rubel2004}

The value of $\tau$ was obtained numerically, as explained in the
previous section with respect to Eq.~(\ref{eq-3DG-A3}).
The calculation of $\tau$ is relatively inexpensive (as compared to Monte
Carlo simulations of the diffusion) and it can be performed over
a wide range of model parameters. For the lattice model used in
Sec.~\ref{sec-simulation}, we have found that the simulated
dependence of $\tau$ on model parameters is well fitted by the
following expressions in two ($\tau_{2D}$) and three ($\tau_{3D}$)
dimensions:
\begin{equation} \label{eq-tau-2d-fit} \tau_{2D} = \left(
    \frac{kT}{\sigma} \right)^{1/4} \exp \left( \lambda_{2D} \left(
      \frac{\sigma}{kT} \right)^{1.7} + \eta_{2D} \right) ,
\end{equation}
\begin{equation} \label{eq-tau-3d-fit}
 \tau_{3D} =  \exp \left( \lambda_{3D} \left( \frac{\sigma}{kT} \right)^{3/2} + \eta_{3D}  \right) ,
\end{equation}
where the coefficients $\lambda$ and $\eta$ depend on the ratio of the
localization radius $a$ to the lattice parameter $d$:
\begin{align}
 \lambda_{2D} &= -1.12 \, (a/d)^{0.2}  + 0.76,  \notag \\
 \eta   _{2D} &=  2.55 \, (d/a)^{0.9}  - 2.4,          \\
 \lambda_{3D} &=  0.27 \, \log (d/a)   - 0.74,  \notag \\
 \eta   _{3D} &=  2.92 \, (d/a)^{0.85} - 3.28.  \notag
\end{align}
The error of fitting does not exceed 2.5\% within the range of
parameters $0.2<kT/\sigma<1$ and  $0.2<a/d<0.5$.

Extensive computer simulations are necessary to determine the
exact value of $\overline{n_{\text{esc}}}$. Without simulations one can
only claim that $\overline{n_{\text{esc}}}$ is a number larger than
unity, though not exponentially large as a function of model
parameters $kT/\sigma$ and $a/d$. Indeed, it follows from the
concept of transport
energy,\cite{Baranovskii1997,Baranovskii2000,Rubel2004} that the
energy of a carrier just after its hop from a trap is close to the
transport energy. It means that there is no energetic barrier for
moving further away from a trap after the first hop. Therefore it
is natural to suppose that, when a carrier has jumped out of a
trap the probability of escaping i.~e.\ 
$(\overline{n_{\text{esc}}})^{-1}$ is comparable to 1.

In order to learn more about $\overline{n_{\text{esc}}}$, we used the values
of $\mu_0$ and $A$ obtained by Monte Carlo simulations.
Substituting the data shown in Fig.~\ref{fig-3D-Tdep} into
Eq.~(\ref{eq-3DG-A3}), we have extracted $\overline{n_{\text{esc}}}$ for
different temperatures in the range $0.33\sigma\leq kT
\leq0.7\sigma$. The localization radius $a$ was chosen equal to
$0.2d$. We have obtained that $\overline{n_{\text{esc}}}$ varies in the
range from $4.5$ to $8$ in the 2D system, and in the range from
$1.6$ to $3.1$ in the 3D system. Simulations have also shown that
the values of $\overline{n_{\text{esc}}}$ do not significantly change
when the localization radius is changed from $0.2d$ to $0.5d$. One can
see that values of $\overline{n_{\text{esc}}}$ found by simulations
are indeed reasonable: they are larger than unity but of that order.

Therefore, the analytical consideration presented in
Sec.~\ref{subsec-hopping-transport} reduces the problem of
predicting the field effect on diffusion to (i) the evaluation of the
zero-field mobility $\mu_0$, and (ii) the evaluation of a coefficient
$\overline{n_{\text{esc}}}$, which is a \emph{slowly varying} function of
the system parameters $kT/\sigma$ and $a/d$. The mobility can be
obtained either from computer simulations (which is a much easier
task than simulations of the diffusion), or from the analytical
theory\cite{Baranovski2006,Rubel2004} and experiments. For the
sake of self-consistency, we rely here on the simulation data for
the mobility. With respect to $\overline{n_{\text{esc}}}$, the comparison
between numerical results and the theory shows that one can consider
this coefficient as a \emph{constant} and nevertheless obtain results
that agree with simulations. The appropriate choice of this
constant for $a=0.2\,d$ is
\begin{equation} \label{eq-n-esc-choice}
 \overline{n_{\text{esc}}} \, \simeq
  \left\{ \begin{array}{l}
  6.0 \quad \mbox{(2D),} \\
  2.3 \quad \mbox{(3D).} \\
  \end{array} \right.
\end{equation}

\begin{figure}
\includegraphics[width=8.6 cm]{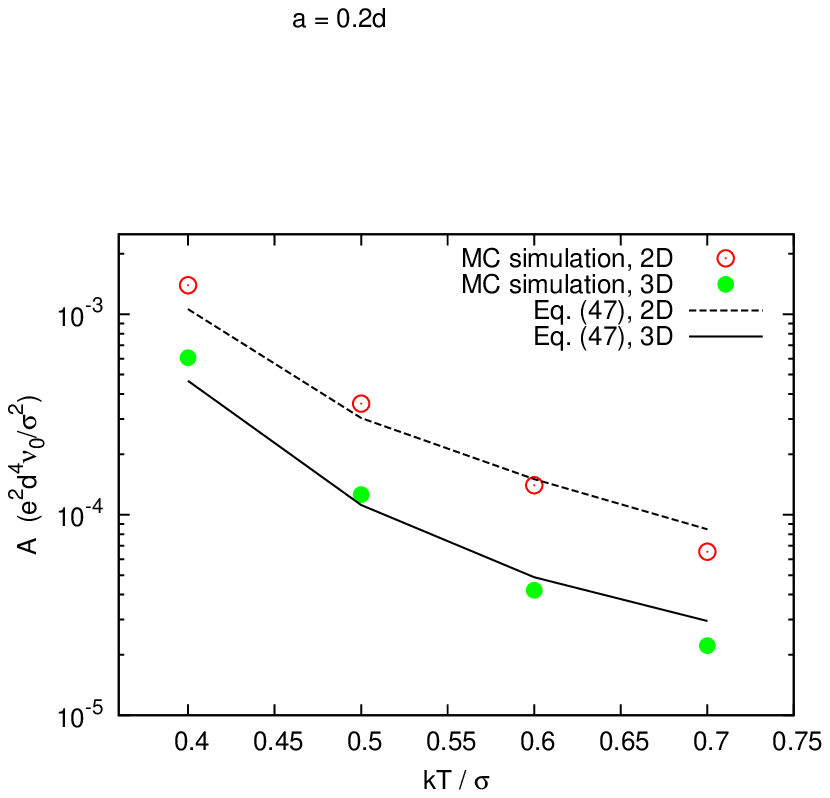}
\caption{Comparison of Monte Carlo results for coefficient $A$
  (symbols) with Eq.~(\ref{eq-3DG-A3}) (lines) for 2D and 3D
  transport. Localization length $a=0.2d$.  Values of $\tau$ and
  $\overline{n_{\text{esc}}}$ are taken from
  Eqs.~(\ref{eq-tau-2d-fit}),(\ref{eq-tau-3d-fit}),(\ref{eq-n-esc-choice}). For
  $\mu_0$ Monte Carlo results are used.}
\label{fig-A_vs_T_02}
\end{figure}

In Fig.~\ref{fig-A_vs_T_02}, Monte Carlo results for the
coefficient $A$, which describes the field-induced diffusion according
to Eq.~(\ref{eq-quadratic-D}), are compared to the analytical
expression~(\ref{eq-3DG-A3}). The coefficient $\overline{n_{\text{esc}}}$
in Eq.~(\ref{eq-3DG-A3}) is set to a constant according to
Eq.~(\ref{eq-n-esc-choice}). It is evident that in the framework
of the simplifying assumption of constant $\overline{n_{\text{esc}}}$,
the analytical theory correctly reproduces the shape of the
temperature dependence for the field-induced diffusion
coefficient.

\section {Conclusions}
\label{sec-conclusions} The main result of this paper is the
development of an analytical theory for the field-induced
diffusion in the hopping transport mode in 3D and 2D systems with
the Gaussian DOS given by Eq.~(\ref{eq-gaussian-dos}). At low
electric fields, the field dependence of the longitudinal diffusion coefficient
$D_z(F)$ is parabolic as expressed in Eq.~(\ref{eq-quadratic-D}).
The analytical expression (\ref{eq-3DG-A3}) gives the temperature
dependence of the field-induced diffusion. Accompanying Monte
Carlo simulations confirm the analytical results and show that the
shape of the field dependence is parabolic.
Together with the
exact results of the previous paper\cite{Nenashev2009I} stating
the non-analytic linear field dependence of the diffusion
coefficient in the 1D case, our result resolves the long-standing
puzzle for the reason of the discrepancies between the analytic
and non-analytic behavior of $D_z(F)$ claimed in different
studies\cite{Bouchaud1989,Pautmeier1991}: It is the space
dimensionality that is responsible for non-analytic (1D) and
analytic (2D and 3D) dependences in $D_z(F)$.

Furthermore, our theory shows that the main contribution to the
field-induced diffusion process comes from localized states with
energies in the vicinity of $-2\sigma^2/kT$. The DOS parameter
$\sigma$ in organic semiconductors is of the order of $0.1$
eV.\cite{Pope1999,Bassler1993,Auweraer1994,Baranovski2006}
Therefore at room temperatures this energy $-2\sigma^2/kT$, which is 
decisive for $D_z(F)$, is situated very deep in the tail of the DOS,
around $-8\sigma$.
This fact raises very severe demands to the size of the system in
computer simulations, which aim at studying the field-induced
diffusion in organic semiconductors.
In order to have the decisive traps in a 
simulation at room temperature one needs approximately $10^{16}$ sites.
Therefore, such simulations cannot be considered
suitable for studying the field-induced diffusion. For
instance, the system size of $70^3$ used in the previous
simulations\cite{Pautmeier1991,Richert1989} is suitable only at
$kT\gtrsim 0.55\sigma$. In contrast, our simulations carried out on
systems with $700^3$ sites give reliable results at
$kT \gtrsim 0.4 \sigma$ confirming the developed analytical theory in a
wide range of parameters.

In all models the influence of the electric field on the
mobility and on the diffusion coefficient increases with
decreasing temperature. For the mobility this phenomenon has been
accounted in the frame of the concept of the effective
temperature.\cite{Jansson2008,Jansson2008b} The results of this
paper show that for the longitudinal diffusion coefficient the
concept of the deep traps should be used instead of the effective
temperature.

\begin{acknowledgments}
Financial support from the Academy of Finland project 116995,
from the Deutsche Forschungsgemeinschaft and
that of the Fonds der Che\-mischen Industrie is gratefully acknowledged.
The calculations were done at the facilities of the Finnish IT
center for science, CSC.
\end{acknowledgments}

\appendix
\section {Monte Carlo algorithm}
\label{sec-MC-details} 
For a numerical study of the diffusion in hopping transport, an
algorithm is needed that can efficiently simulate transport in large
systems. Since the number of sites that can be treated in the
simulation is limited by the available memory, we have chosen to study
hopping transport on a lattice instead of a system with randomly
placed sites.

When the charge carrier is located at site $i$, the probability that
the next jump takes it to the site $j$ is given by
\begin {equation}
p_j = \frac{\Gamma_{ij}}{\Gamma_i},
\end{equation}
where $\Gamma_i$ is the total rate of hopping away from site $i$:
\begin {equation}
\Gamma_i = \sum_j \Gamma_{ij}.
\end{equation}
The time $\tau$ that the charge carrier spends on the site $i$ before hopping,
(the ``dwell time'') is calculated as
\begin {equation}
\tau = T/\Gamma_i,
\end {equation}
where $T$ for each hop is randomly generated with an exponential distribution
with unit variance.
Which jump to perform is decided by picking a random number $x$ between 0 and 1
from a uniform distribution,
and finding $j$ such that
\begin {equation}
\label{eq-MC-choise}
\sum_{k=1}^{j-1} p_k \le x < \sum_{k=1}^{j} p_k
\end {equation}
This ensures that each site $k$ is selected with probability $p_k$.
So far this is the standard Monte Carlo algorithm for hopping
transport.\cite{Richert1989,Bassler1993,Pautmeier1991} Below an efficient
implementation of this algorithm will be described.

Calculating the hopping rates (\ref{eq-Miller-Abrahams}) is very time
consuming since the exponential function is expensive to compute.  
If we instead of the
site energies $\varepsilon_i$ store their exponentials,
\begin {equation}
\kappa_i = \exp \left ( -\frac{\varepsilon_i} {kT} \right),
\end {equation}
the hopping rates $\Gamma_{ij}$ can be computed more efficiently.  We use the
fact that only a small number of discrete displacements are possible
for hopping in a lattice, when we restrict the length of the hops.  
Therefore the geometric part of the hopping rate and the energy
contribution from the electric field can be calculated and stored 
once for each displacement.  
In our case the hops are restricted to a cube of $7 \times 7 \times 7$
sites centered at the starting site.
For each displacement $\Delta r$,
define the quantities
\begin {equation}
\rho_{\Delta r} = \exp \left( -2 \frac{\Delta r} {a} \right)
\end {equation}
and
\begin {equation}
\varphi_{\Delta r} = \exp \left( \frac{eF \Delta z}{kT} \right),
\end {equation}
where $\Delta z$ is the $z$-component of $\Delta r$.
The hopping rate from site $i$ to site $j$ located at
the position $\Delta r$ relative to $i$ can now be evaluated using
\begin {equation}
  \label{eq-MC-fastrate}
  \Gamma_{ij} = \nu_0 \rho_{\Delta r} \min \left( 1, \frac{\kappa_j}{\kappa_i}
  \varphi_{\Delta r} \right).
\end {equation}

One could also consider the storing of all hopping rates $\Gamma_{ij}$, 
but since this greatly increases the amount of memory needed by the
simulation, it would restrict the size of the systems that can be
simulated.  We have found that a good balance between the simulation speed
and the memory requirements is achieved by storing only the rate
$\Gamma_i$ and the quantity $\kappa_i$ for each site, and to calculate
the rates $\Gamma_{ij}$ by Eq.~(\ref{eq-MC-fastrate}) during the 
simulation each time they are needed. 

One further optimization is to consider the jumps in order of
increasing lengths, since shorter jumps typically have higher
probabilities than longer ones. This ordering greatly reduces the
number of hopping rates that have to be evaluated before the
destination site $j$ that satisfies Eq.\ (\ref{eq-MC-choise}) is found.

Each carrier was initially placed on a randomly chosen site in the
lattice and then allowed to equilibrate before the measurements of
time and displacements started.

\newpage


\begin{thebibliography}{26}
\expandafter\ifx\csname natexlab\endcsname\relax\def\natexlab#1{#1}\fi
\expandafter\ifx\csname bibnamefont\endcsname\relax
  \def\bibnamefont#1{#1}\fi
\expandafter\ifx\csname bibfnamefont\endcsname\relax
  \def\bibfnamefont#1{#1}\fi
\expandafter\ifx\csname citenamefont\endcsname\relax
  \def\citenamefont#1{#1}\fi
\expandafter\ifx\csname url\endcsname\relax
  \def\url#1{\texttt{#1}}\fi
\expandafter\ifx\csname urlprefix\endcsname\relax\def\urlprefix{URL }\fi
\providecommand{\bibinfo}[2]{#2}
\providecommand{\eprint}[2][]{\url{#2}}

\bibitem[{\citenamefont{Baranovski}(2006)}]{Baranovski2006}
\bibinfo{editor}{\bibfnamefont{S.}~\bibnamefont{Baranovski}}, ed.,
  \emph{\bibinfo{title}{Charge Transport in Disordered Solids with Applications
  in Electronics}} (\bibinfo{publisher}{John Wiley \& Sons, Ltd, Chichester},
  \bibinfo{year}{2006}).

\bibitem[{\citenamefont{Roth}(1991)}]{Roth1991}
\bibinfo{author}{\bibfnamefont{S.}~\bibnamefont{Roth}}, in
  \emph{\bibinfo{booktitle}{Hopping transport in solids}}, edited by
  \bibinfo{editor}{\bibfnamefont{M.}~\bibnamefont{Pollak}} \bibnamefont{and}
  \bibinfo{editor}{\bibfnamefont{B.~I.} \bibnamefont{Shklovskii}}
  (\bibinfo{publisher}{Elsevier}, \bibinfo{year}{1991}), p.
  \bibinfo{pages}{377}.

\bibitem[{\citenamefont{B\"{a}ssler}(2000)}]{Bassler2000}
\bibinfo{author}{\bibfnamefont{H.}~\bibnamefont{B\"{a}ssler}},
  \emph{\bibinfo{title}{Semiconducting Polymers, G. Hadziioannou and P. F. van
  Hutten (eds.)}} (\bibinfo{publisher}{John Wiley \& Sons, Inc., New York},
  \bibinfo{year}{2000}), p. \bibinfo{pages}{365}.

\bibitem[{\citenamefont{Pope and Swenberg}(1999)}]{Pope1999}
\bibinfo{author}{\bibfnamefont{M.}~\bibnamefont{Pope}} \bibnamefont{and}
  \bibinfo{author}{\bibfnamefont{C.~E.} \bibnamefont{Swenberg}},
  \emph{\bibinfo{title}{Electronic Processes in Organic Crystals and Polymers}}
  (\bibinfo{publisher}{Oxford University Press, Oxford}, \bibinfo{year}{1999}).

\bibitem[{\citenamefont{B\"assler}(1993)}]{Bassler1993}
\bibinfo{author}{\bibfnamefont{H.}~\bibnamefont{B\"assler}},
  \bibinfo{journal}{Phys. Status Solidi B} \textbf{\bibinfo{volume}{175}},
  \bibinfo{pages}{15} (\bibinfo{year}{1993}).

\bibitem[{\citenamefont{Borsenberger
  et~al.}(1993{\natexlab{a}})\citenamefont{Borsenberger, Magin, van~der
  Auweraer, and de~Schryver}}]{Borsenberger1993}
\bibinfo{author}{\bibfnamefont{P.~M.} \bibnamefont{Borsenberger}},
  \bibinfo{author}{\bibfnamefont{E.~H.} \bibnamefont{Magin}},
  \bibinfo{author}{\bibfnamefont{M.}~\bibnamefont{van~der Auweraer}},
  \bibnamefont{and} \bibinfo{author}{\bibfnamefont{F.~C.}
  \bibnamefont{de~Schryver}}, \bibinfo{journal}{Phys. Status Solidi A}
  \textbf{\bibinfo{volume}{140}}, \bibinfo{pages}{9}
  (\bibinfo{year}{1993}{\natexlab{a}}).

\bibitem[{\citenamefont{van~der Auweraer et~al.}(1994)\citenamefont{van~der
  Auweraer, de~Schryver, Borsenberger, and B\"assler}}]{Auweraer1994}
\bibinfo{author}{\bibfnamefont{M.}~\bibnamefont{van~der Auweraer}},
  \bibinfo{author}{\bibfnamefont{F.~C.} \bibnamefont{de~Schryver}},
  \bibinfo{author}{\bibfnamefont{P.~M.} \bibnamefont{Borsenberger}},
  \bibnamefont{and}
  \bibinfo{author}{\bibfnamefont{H.}~\bibnamefont{B\"assler}},
  \bibinfo{journal}{Advanced Materials} \textbf{\bibinfo{volume}{6}},
  \bibinfo{pages}{199} (\bibinfo{year}{1994}).

\bibitem[{\citenamefont{Nenashev et~al.}(2009)\citenamefont{Nenashev, Jansson,
  Baranovskii, \"Oster\-backa, Dvurechenskii, and Gebhard}}]{Nenashev2009I}
\bibinfo{author}{\bibfnamefont{A.~V.} \bibnamefont{Nenashev}},
  \bibinfo{author}{\bibfnamefont{F.}~\bibnamefont{Jansson}},
  \bibinfo{author}{\bibfnamefont{S.~D.} \bibnamefont{Baranovskii}},
  \bibinfo{author}{\bibfnamefont{R.}~\bibnamefont{\"Oster\-backa}},
  \bibinfo{author}{\bibfnamefont{A.~V.} \bibnamefont{Dvurechenskii}},
  \bibnamefont{and} \bibinfo{author}{\bibfnamefont{F.}~\bibnamefont{Gebhard}},
  \bibinfo{journal}{arXiv:0912.3161}  (\bibinfo{year}{2009}).

\bibitem[{\citenamefont{{H.-J. Yuh} and Stolka}(1988)}]{Yuh1988}
\bibinfo{author}{\bibnamefont{{H.-J. Yuh}}} \bibnamefont{and}
  \bibinfo{author}{\bibfnamefont{M.}~\bibnamefont{Stolka}},
  \bibinfo{journal}{Philos. Mag. B} \textbf{\bibinfo{volume}{58}},
  \bibinfo{pages}{539} (\bibinfo{year}{1988}).

\bibitem[{\citenamefont{Borsenberger et~al.}(1991)\citenamefont{Borsenberger,
  Pautmeier, Richert, and B\"assler}}]{Borsenberger1991}
\bibinfo{author}{\bibfnamefont{P.~M.} \bibnamefont{Borsenberger}},
  \bibinfo{author}{\bibfnamefont{L.}~\bibnamefont{Pautmeier}},
  \bibinfo{author}{\bibfnamefont{R.}~\bibnamefont{Richert}}, \bibnamefont{and}
  \bibinfo{author}{\bibfnamefont{H.}~\bibnamefont{B\"assler}},
  \bibinfo{journal}{Journal of Chemical Physics} \textbf{\bibinfo{volume}{94}},
  \bibinfo{pages}{8276} (\bibinfo{year}{1991}).

\bibitem[{\citenamefont{Borsenberger
  et~al.}(1993{\natexlab{b}})\citenamefont{Borsenberger, Richert, and
  B\"assler}}]{Borsenberger1993b}
\bibinfo{author}{\bibfnamefont{P.~M.} \bibnamefont{Borsenberger}},
  \bibinfo{author}{\bibfnamefont{R.}~\bibnamefont{Richert}}, \bibnamefont{and}
  \bibinfo{author}{\bibfnamefont{H.}~\bibnamefont{B\"assler}},
  \bibinfo{journal}{Phys. Rev. B} \textbf{\bibinfo{volume}{47}},
  \bibinfo{pages}{4289} (\bibinfo{year}{1993}{\natexlab{b}}).

\bibitem[{\citenamefont{Hirao and Nishizawa}(1996)}]{Hirao1996}
\bibinfo{author}{\bibfnamefont{A.}~\bibnamefont{Hirao}} \bibnamefont{and}
  \bibinfo{author}{\bibfnamefont{H.}~\bibnamefont{Nishizawa}},
  \bibinfo{journal}{Phys. Rev. B} \textbf{\bibinfo{volume}{54}},
  \bibinfo{pages}{4755} (\bibinfo{year}{1996}).

\bibitem[{\citenamefont{Hirao and Nishizawa}(1997)}]{Hirao1997}
\bibinfo{author}{\bibfnamefont{A.}~\bibnamefont{Hirao}} \bibnamefont{and}
  \bibinfo{author}{\bibfnamefont{H.}~\bibnamefont{Nishizawa}},
  \bibinfo{journal}{Phys. Rev. B} \textbf{\bibinfo{volume}{56}},
  \bibinfo{pages}{R2904} (\bibinfo{year}{1997}).

\bibitem[{\citenamefont{Lupton and Klein}(2002)}]{Lupton2002}
\bibinfo{author}{\bibfnamefont{J.~M.} \bibnamefont{Lupton}} \bibnamefont{and}
  \bibinfo{author}{\bibfnamefont{J.}~\bibnamefont{Klein}},
  \bibinfo{journal}{Phys. Rev. B} \textbf{\bibinfo{volume}{65}},
  \bibinfo{pages}{193202} (\bibinfo{year}{2002}).

\bibitem[{\citenamefont{Harada et~al.}(2005)\citenamefont{Harada, Werner,
  Pfeiffer, Bloom, Elliott, and Leo}}]{Harada2005}
\bibinfo{author}{\bibfnamefont{K.}~\bibnamefont{Harada}},
  \bibinfo{author}{\bibfnamefont{A.~G.} \bibnamefont{Werner}},
  \bibinfo{author}{\bibfnamefont{M.}~\bibnamefont{Pfeiffer}},
  \bibinfo{author}{\bibfnamefont{C.~J.} \bibnamefont{Bloom}},
  \bibinfo{author}{\bibfnamefont{C.~M.} \bibnamefont{Elliott}},
  \bibnamefont{and} \bibinfo{author}{\bibfnamefont{K.}~\bibnamefont{Leo}},
  \bibinfo{journal}{Phys. Rev. Lett.} \textbf{\bibinfo{volume}{94}},
  \bibinfo{pages}{036601} (\bibinfo{year}{2005}).

\bibitem[{\citenamefont{Miller and Abrahams}(1960)}]{Miller1960}
\bibinfo{author}{\bibfnamefont{A.}~\bibnamefont{Miller}} \bibnamefont{and}
  \bibinfo{author}{\bibfnamefont{E.}~\bibnamefont{Abrahams}},
  \bibinfo{journal}{Phys. Rev.} \textbf{\bibinfo{volume}{120}},
  \bibinfo{pages}{745} (\bibinfo{year}{1960}).

\bibitem[{\citenamefont{Richert et~al.}(1989)\citenamefont{Richert, Pautmeier,
  and B\"assler}}]{Richert1989}
\bibinfo{author}{\bibfnamefont{R.}~\bibnamefont{Richert}},
  \bibinfo{author}{\bibfnamefont{L.}~\bibnamefont{Pautmeier}},
  \bibnamefont{and}
  \bibinfo{author}{\bibfnamefont{H.}~\bibnamefont{B\"assler}},
  \bibinfo{journal}{Phys. Rev. Lett.} \textbf{\bibinfo{volume}{63}},
  \bibinfo{pages}{547} (\bibinfo{year}{1989}).

\bibitem[{\citenamefont{Pautmeier et~al.}(1991)\citenamefont{Pautmeier,
  Richert, and B\"assler}}]{Pautmeier1991}
\bibinfo{author}{\bibfnamefont{L.}~\bibnamefont{Pautmeier}},
  \bibinfo{author}{\bibfnamefont{R.}~\bibnamefont{Richert}}, \bibnamefont{and}
  \bibinfo{author}{\bibfnamefont{H.}~\bibnamefont{B\"assler}},
  \bibinfo{journal}{Philos. Mag. B} \textbf{\bibinfo{volume}{63}},
  \bibinfo{pages}{587} (\bibinfo{year}{1991}).

\bibitem[{\citenamefont{Orenstein and Kastner}(1981)}]{Orenstein1981}
\bibinfo{author}{\bibfnamefont{J.}~\bibnamefont{Orenstein}} \bibnamefont{and}
  \bibinfo{author}{\bibfnamefont{M.}~\bibnamefont{Kastner}},
  \bibinfo{journal}{Solid State Commun.} \textbf{\bibinfo{volume}{40}},
  \bibinfo{pages}{85} (\bibinfo{year}{1981}).

\bibitem[{\citenamefont{Baranovskii et~al.}(1997)\citenamefont{Baranovskii,
  Faber, Hensel, and Thomas}}]{Baranovskii1997}
\bibinfo{author}{\bibfnamefont{S.~D.} \bibnamefont{Baranovskii}},
  \bibinfo{author}{\bibfnamefont{T.}~\bibnamefont{Faber}},
  \bibinfo{author}{\bibfnamefont{F.}~\bibnamefont{Hensel}}, \bibnamefont{and}
  \bibinfo{author}{\bibfnamefont{P.}~\bibnamefont{Thomas}},
  \bibinfo{journal}{J. Phys. C} \textbf{\bibinfo{volume}{9}},
  \bibinfo{pages}{2699} (\bibinfo{year}{1997}).

\bibitem[{\citenamefont{Baranovskii et~al.}(2000)\citenamefont{Baranovskii,
  Cordes, Hensel, and Leising}}]{Baranovskii2000}
\bibinfo{author}{\bibfnamefont{S.~D.} \bibnamefont{Baranovskii}},
  \bibinfo{author}{\bibfnamefont{H.}~\bibnamefont{Cordes}},
  \bibinfo{author}{\bibfnamefont{F.}~\bibnamefont{Hensel}}, \bibnamefont{and}
  \bibinfo{author}{\bibfnamefont{G.}~\bibnamefont{Leising}},
  \bibinfo{journal}{Phys. Rev. B} \textbf{\bibinfo{volume}{62}},
  \bibinfo{pages}{7934} (\bibinfo{year}{2000}).

\bibitem[{\citenamefont{Rubel et~al.}(2004)\citenamefont{Rubel, Baranovskii,
  Thomas, and Yamasaki}}]{Rubel2004}
\bibinfo{author}{\bibfnamefont{O.}~\bibnamefont{Rubel}},
  \bibinfo{author}{\bibfnamefont{S.~D.} \bibnamefont{Baranovskii}},
  \bibinfo{author}{\bibfnamefont{P.}~\bibnamefont{Thomas}}, \bibnamefont{and}
  \bibinfo{author}{\bibfnamefont{S.}~\bibnamefont{Yamasaki}},
  \bibinfo{journal}{Phys. Rev. B} \textbf{\bibinfo{volume}{69}},
  \bibinfo{pages}{014206} (\bibinfo{year}{2004}).

\bibitem[{\citenamefont{Rudenko and Arkhipov}(1982)}]{Rudenko1982}
\bibinfo{author}{\bibfnamefont{A.~I.} \bibnamefont{Rudenko}} \bibnamefont{and}
  \bibinfo{author}{\bibfnamefont{V.~I.} \bibnamefont{Arkhipov}},
  \bibinfo{journal}{Philos. Mag. B} \textbf{\bibinfo{volume}{45}},
  \bibinfo{pages}{177} (\bibinfo{year}{1982}).

\bibitem[{\citenamefont{Bouchaud and Georges}(1989)}]{Bouchaud1989}
\bibinfo{author}{\bibfnamefont{J.~P.} \bibnamefont{Bouchaud}} \bibnamefont{and}
  \bibinfo{author}{\bibfnamefont{A.}~\bibnamefont{Georges}},
  \bibinfo{journal}{Phys. Rev. Lett.} \textbf{\bibinfo{volume}{63}},
  \bibinfo{pages}{2692} (\bibinfo{year}{1989}).

\bibitem[{\citenamefont{Jansson
  et~al.}(2008{\natexlab{a}})\citenamefont{Jansson, Baranovskii, Sliau\v{z}ys,
  \"Osterbacka, and Thomas}}]{Jansson2008}
\bibinfo{author}{\bibfnamefont{F.}~\bibnamefont{Jansson}},
  \bibinfo{author}{\bibfnamefont{S.~D.} \bibnamefont{Baranovskii}},
  \bibinfo{author}{\bibfnamefont{G.}~\bibnamefont{Sliau\v{z}ys}},
  \bibinfo{author}{\bibfnamefont{R.}~\bibnamefont{\"Osterbacka}},
  \bibnamefont{and} \bibinfo{author}{\bibfnamefont{P.}~\bibnamefont{Thomas}},
  \bibinfo{journal}{Phys. Status Solidi C} \textbf{\bibinfo{volume}{5}},
  \bibinfo{pages}{722} (\bibinfo{year}{2008}{\natexlab{a}}).

\bibitem[{\citenamefont{Jansson
  et~al.}(2008{\natexlab{b}})\citenamefont{Jansson, Baranovskii, Gebhard, and
  \"{O}sterbacka}}]{Jansson2008b}
\bibinfo{author}{\bibfnamefont{F.}~\bibnamefont{Jansson}},
  \bibinfo{author}{\bibfnamefont{S.~D.} \bibnamefont{Baranovskii}},
  \bibinfo{author}{\bibfnamefont{F.}~\bibnamefont{Gebhard}}, \bibnamefont{and}
  \bibinfo{author}{\bibfnamefont{R.}~\bibnamefont{\"{O}sterbacka}},
  \bibinfo{journal}{Phys. Rev. B} \textbf{\bibinfo{volume}{77}},
  \bibinfo{eid}{195211} (\bibinfo{year}{2008}{\natexlab{b}}).

\end{thebibliography}

\end{document}